# Impact of TCO Microstructure on the Electronic Properties of Carbazole-based Self-Assembled Monolayers

*Suzana Kralj [1a]\*, Pia Dally [2], Pantelis Bampoulis [1b], Badri Vishal [2], Stefaan De Wolf [2], Monica Morales-Masis [1a]\**

AUTHOR ADDRESS

[1a] S. Kralj, M. Morales-Masis

MESA+ Institute for Nanotechnology, University of Twente, Enschede 7500 AE, The Netherlands

[1b] P. Bampoulis

Physics of Interfaces and Nanomaterials, MESA+ Institute for Nanotechnology, University of Twente, Enschede 7500 AE, The Netherlands

[2] P. Dally, B. Vishal, S. De Wolf

KAUST Solar Center (KSC), Physical Sciences and Engineering Division (PSE), King Abdullah University of Science and Technology (KAUST), Thuwal 23955-6900, Kingdom of Saudi Arabia

AUTHOR INFORMATION

**Corresponding Author**

\* S. Kralj (s.k.kralj@utwente.nl), \* M. Morales-Masis (m.moralesmasis@utwente.nl)

ABSTRACT

Carbazole-based self-assembled monolayers (PACz-SAMs), anchored via their phosphonic acid group on a transparent conductive oxide (TCO) have demonstrated excellent performance as hole-selective layers in inverted perovskite solar cells. However, the influence of the TCO microstructure on the work function (WF) shift after SAM anchoring as well as the WF variations at the micro/nanoscale have not been extensively studied yet. Herein, we investigate the effect of the Sn-doped $In_2O_3$ (ITO) microstructure on the WF distribution upon 2PACz-SAMs and $NiO_x$/2PACz-SAMs application. For this, ITO substrates with amorphous and polycrystalline (featuring either nanoscale or microscale-sized grains) microstructures are studied. A correlation between the ITO grain orientation and 2PACz-SAMs local potential distribution was found via Kelvin probe force microscopy and electron backscatter diffraction. These variations vanish for amorphous ITO or when adding an amorphous $NiO_x$ buffer layer, where a homogeneous surface potential distribution is mapped. Ultraviolet photoelectron spectroscopy confirmed the ITO WF increase after 2PACz-SAMs deposition. Considering the importance of polycrystalline TCOs as high mobility and broadband transparent electrodes, we provide insights to ensure uniform WF distribution upon application of hole transport SAMs, which is critical towards enhanced device performance.

MAIN TEXT

Transparent conductive oxides (TCOs) are omnipresent in a range of high-efficiency optoelectronic devices, including perovskite solar cells (PSCs), both in their single junction and tandem implementations[1,2]. Among the various available TCOs, indium tin oxide (ITO) remains a common choice as a transparent electrode for optoelectronic applications as it is well established, scalable, and commercial glass/ITO substrates are readily available. However, further progress in PSC performance could benefit from additional TCO optimization. Moreover, with the rise of monolithic perovskite/perovskite and perovskite/silicon tandem solar cells, TCOs are also often used as an interband recombination junction[3] deposited onto the bottom cell (i.e., a perovskite or silicon cell), connecting the subcells in series. Both in the single-junction and in the tandem case for inverted, *p-i-n* devices, the hole transport layer (HTL) is deposited onto the TCO, followed by the perovskite absorber and the electron selective contact stack. Depending on the TCO deposition method and process conditions, different microstructures can be achieved which may also influence the optoelectronic properties. For instance, amorphous TCOs generally feature a narrower band gap as compared to polycrystalline TCOs due to their distorted absorption edge[4,5]. The electron mobility in TCOs can also be influenced by the microstructure, with typically high mobilities achieved for polycrystalline TCOs with large grains[5-9].

For the case of HTLs for inverted PSCs, self-assembled monolayers (SAMs) such as carbazole-based with a phosphonic acid anchoring group (PACz-SAMs), have attracted much attention in recent years[10-12]. These SAMs shift the work function of the TCO substrate to higher values due to the interfacial dipole they introduce, enhancing the hole selectivity of the contact[12-14]. So far, the TCO/PACz-SAMs stack has gained particular attention as a hole selective contact for inverted perovskite[15-21], organic[22,23], perovskite/organic[24], perovskite/perovskite[25-27], and recent record perovskite/silicon tandem solar cells[28, 29] where often they have found to result in superior passivation of the HTL/photoabsorber interface, a fast hole extraction rate and minimal parasitic absorption[30]. On the downside, the presence of imperfections in PACz-SAMs coatings on polycrystalline TCO is frequently reported,[30-33] preventing optimal device performance and stability. So far, addressing this challenge has primarily involved either blending SAMs molecules of varying sizes to ensure high packing density[26, 33-36] or anchoring the PACz-SAMs to a hole-selective metal-oxide buffer layer, such as nickel oxide ($NiO_x$)[37-40]. However, to the best of our knowledge, no comprehensive investigation into the correlation between the microstructure of TCOs and the hole transporting properties of TCO/PACz-SAMs stacks has been reported. To elucidate this, herein we study the effect of 2PACz, [2-(9H-carbazol-9-yl)ethyl]phosphonic acid (**Figure S1**), a commonly used SAM hole selective contact, on the WF shifts of ITO substrates. Furthermore, we investigate the potential distribution and its link to the WF along the surface of various types of ITO substrates with different microstructures. We focus in this work on ITO as a model system, but the drawn conclusions are valid for other TCOs. Specifically, we studied ITO substrates with comparable sheet resistance but distinct microstructures, namely: commercial ITO, featuring a polycrystalline microstructure with small (nm-scale) grains, and pulsed laser deposited (PLD) ITO, either amorphous or polycrystalline with large (μm-scale) grains. Moreover, the effect of introducing a sputtered $NiO_x$ layer between the different ITO electrodes and 2PACz-SAMs was analysed. The potential distribution was mapped using Kelvin probe force microscopy (KPFM), while the grain orientation of the ITO for the same area was measured by electron backscatter diffraction (EBSD) analysis. Ultraviolet photoelectron spectroscopy (UPS) was used to determine absolute WF values and verify the values determined by KPFM. Based on these experiments, we demonstrate how the ITO crystalline grain orientation and grain size influence the potential distribution in the ITO/2PACz-SAMs electrodes, compare the WF values achieved for the distinct ITOs and discuss the role of a NiO$_x$ buffer layer on achieving a uniform potential distribution for hole extraction. By optimizing the TCO/PACz-SAMs interface, valuable insights for enhancing solar cell designs can be gained, thus further improving their efficiency and reliability.

EXPERIMENTAL SECTION

100 nm Sn-doped In$_2$O$_3$ (10/90 wt% SnO$_2$/In$_2$O$_3$) thin-films with sheet resistance ($R_{sh}$) below 50 Ω/sq, representing typical TCO device requirements, with three different microstructures were selected for this study. The ITO films were deposited on glass substrates by PLD at room temperature. As-deposited ITO films were found to be amorphous, subsequent annealing for 20 min at 450˚C (**Figure S2**), resulted in a polycrystalline structure, as confirmed by XRD (**Figure 1a, Table S1**). Top-view SEM and AFM scans (**Figure 1c-d**) show a flat, featureless surface for the amorphous ITO (RMS of 0.29 nm), and large micron-sized grains for the annealed ITO films (RMS of 0.30 nm). This observed change in crystallinity and microstructure is due to a solid phase crystallization, as previously reported for sputtered In-based TCOs[7, 8]. In the process of physical vapor deposition of In-based TCOs at room temperature (either by PLD or sputtering), nanocrystals are generated within an amorphous matrix. These nanocrystals act as nucleation sites, facilitating the growth of grains during a subsequent annealing step[7]. Commercially available ITO substrates (Ossila Ltd.), featuring a polycrystalline structure and nano-scale grains microstructure (RMS of 3.20 nm), as depicted in **Figure 1a** and **e**, respectively, were used to compare the influence of the grain size in the WF distribution. Furthermore, all studied ITO films demonstrate >80% transmittance in the wavelength range of 350–750 nm (**Figure 1b**). For wavelengths above 750 nm, the commercial ITO samples exhibited a larger absorption (>15%) as compared to PLD ITO. This can be explained by free carrier absorption as the concentration of free carriers, $N_e$, is significantly higher (10x10$^{20}$ cm$^{-3}$) for commercial ITO films as compared to PLD films (up to 4.5x10$^{20}$ cm$^{-3}$). The electrical properties of the ITO films are summarized in the inset of **Figure 1b**. Additionally, an increase in the optical band gap for PLD ITO annealed films from ~3.4 eV for as deposited ITO to ~3.7 eV for annealed ITO was estimated from the Tauc plot in **Figure S3.**).

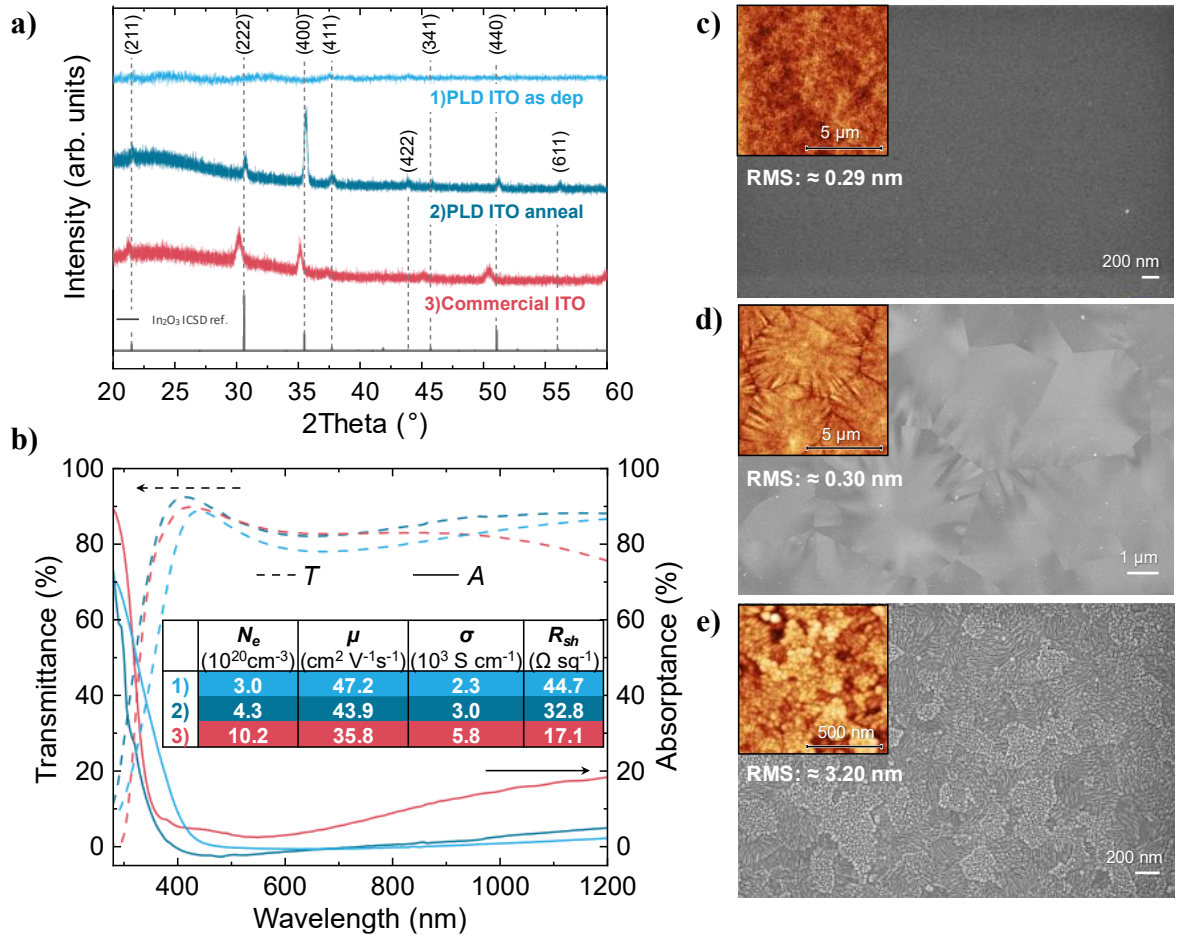

**Figure 1**. **(a)** X-ray diffractogram; and **(b)** Optical and electrical properties of studied ITOs. SEM top-view images with AFM topography inset and RMS values for **(c)** PLD ITO as deposited; **(d)** PLD ITO annealed; **(e)** Commercial ITO. The thickness of all films is ≈100 nm.

For conciseness, as-deposited ITO films will further be referred as '*a-ITO*', annealed ITO polycrystalline films with large (µm-sized) grains as '*poly-ITO-µm-grain*' and commercial ITO polycrystalline films with small (nm-sized) grains as '*poly-ITO-nm-grain*'.

To gain insight into the surface potential distribution of the ITO films with distinct microstructural properties, KPFM was performed, which directly maps the contact potential difference (CPD) between a conducting AFM tip and the sample (**Figure S4**). Here, a relatively large area of 10x10 µm$^2$ was scanned. The resulting CPD maps and the corresponding topography image (inset) are shown in the left column in **Figure 2a**. In the case of *a-ITO* and *poly-ITO-nm-grains*, a relatively consistent surface potential across the scanned film area is measured, indicating an overall homogeneous WF distribution. However, in the case of *poly-ITO-µm-grain* films, the CPD is not uniform across the scanned area, as observed by the lighter and darker domains in **Figure 2a**, representing areas of higher and lower WF values, respectively. The domains match with the corresponding large grains in the topography inset as marked with white arrows.

Subsequently, 2PACz-SAMs were deposited onto the three ITO films described above (details in SI). The right column in **Figure 2a** displays the CPD maps, accompanied by insets of topography AFM images of the ITO/2PACz-SAMs substrates. Generally, the topography remained virtually unaltered with the introduction of 2PACz-SAMs, but a systematic overall reduction in CPD confirms the presence of 2PACz-SAMs on the surface and implies an increase in the WF. Notably, in the case of the *poly-ITO-µm-grain* films, the presence of domains with distinct CPD values remain even after the 2PACz-SAMs application.

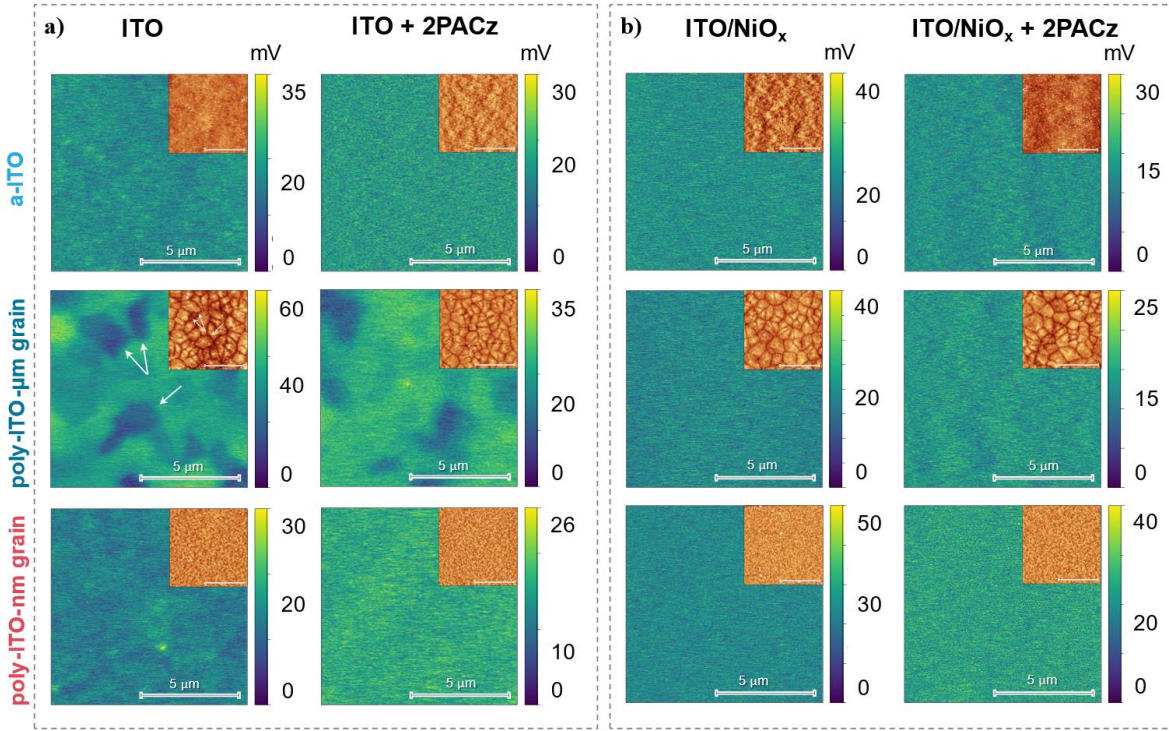

**Figure 2.** KPFM mapping for the stacks: **(a)** ITO with and without 2PACz-SAMs; **(b)** ITO/NiO$_x$ with and without 2PACz-SAMs (scanning area 10x10 μm$^2$, larger image: CPD mapping, inset: topography. Note: To ensure a precise depiction of CPD values for each sample, the colour scale bar has been adjusted to accommodate the observed deviations. The scale bar of all images, including insets, are equivalent to 5 μm).

NiO$_x$ is a high work function metal oxide and has been demonstrated as an effective hole transport layer in inverted PSCs[41, 42]. However, it is widely reported that its direct contact with the perovskite absorber, leads to a defective interface, leading to the development of several NiO$_x$ surface passivation approaches, including the use of PACz-SAMs[37, 43, 44]. While similar performance have been reported for PSCs with ITO/2PACz-SAMs and ITO/NiO$_x$/2PACz-SAMs, it has been proposed[40, 44] that the use of a thin NiO$_x$ buffer layer deposited on top of ITO helps to homogenize morphological and energetical differences on the ITO substrates, enabling higher reproducibility in devices for the ITO/NiO$_x$/2PACz-SAMs stack as compared to the ITO/2PACz-SAMs counterpart[37, 39]. It is hypothesized that the presence of a NiO$_x$ buffer, could minimize the impact of the presence of pin holes in the 2PACz-SAMs layer[38, 40]. Moreover, most of the reported NiO$_x$ hole transport layers, are either amorphous or nanocrystalline, with randomly oriented nanometre-sized grains. We suggest that this amorphous or nanocrystalline microstructure is also beneficial for homogenizing the surface roughness and the surface potential as will be elaborated below.

To confirm this, here we sputtered an amorphous NiO$_x$ layer (14 nm) onto the studied ITOs. **Figure S5** displays a featureless X-ray diffraction pattern, confirming the formation of an amorphous NiO$_x$ film. For elaborate characterizations and properties of the NiO$_x$ layer, we refer the reader to reference[41]. We note that NiO$_x$ subsequently underwent a treatment with a potassium chloride (KCl) solution to passivate its surface defects[45]. Topography and surface potential images for the resulting ITO/NiO$_x$ stack with and without 2PACz-SAMs are shown in **Figure 2b**. The left-hand column illustrates CPD mappings and topography insets for the ITO/NiO$_x$ configuration. AFM images reveal that the NiO$_x$ layer follows the topological features of the distinct ITO substrates. However, the presence of NiO$_x$ on the surface of the ITO films clearly reduces the variations in CPD values along the surface of all ITOs. Consequently, a uniform surface potential distribution emerges, regardless of the underlying microstructure of the ITO. This underscores the efficacy of utilizing an amorphous NiO$_x$ buffer layer as a surface modifier, effectively countering potential non-uniformities and promoting uniform electrical response in the films, particularly in the case of *poly-ITO-µm grain* films. Following the KCl-treated NiO$_x$, 2PACz-SAMs were deposited on ITO/NiO$_x$ substrates (**Figure 2b**, right column). With the incorporation of 2PACz-SAMs, the uniform CPD was preserved for all the studied ITOs/NiO$_x$, but a reduction in CPD values serves as evidence of the presence of 2PACz on the surface, also indicating a WF increase. The WF values were later confirmed by UPS (**Figure *4***).

**Co-localization KPFM and EBSD mapping: correlating ITO grain orientation with work function**

The KPFM findings reveal a notable correspondence between the CPD domains and the respective ITO grains in the *poly-ITO-µm-grain* films. To ascertain whether the identified domains originate from distinct crystalline orientations, a combined approach employing EBSD mapping and KPFM measurements is adopted for simultaneous topographical, electronic and microstructural imaging of the ITO surface on the same point of interest (POI). This is schematically represented in **Figure *3*a** (details on protocol procedure[46] in SI).

It is well-established that crystal orientations can lead to diverse atomic arrangements on a material's surface, influencing the electronic structure and impacting the WF. It has been proposed that generally, closely packed planes (high atomic density) display higher WF compared to loosely packed planes (low atomic density)[47]. From reported surface densities based on density functional theory calculations for $In_2O_3$ (100) > (110) > (111), it is speculated that ITO (100) planes possess higher WF[48]. Another theory that may elucidate the phenomenon of crystalline orientation's influence on local WF variation is the surface polarity concept, introduced by Tasker[49]. While the initial observations were for $In_2O_3$[48, 50, 51] this concept can be extended to ITO. In detail, (111) plane is a Tasker II type of surface without surface dipole perpendicular to the surface normal. On the other hand, (100) plane corresponds to a Tasker III type of surface, characterized by alternating charged planes that lead to a dipolar moment on the uppermost surface layer pointing away from the surface in the normal direction. Consequently, this induced surface dipole makes the removal of an electron more challenging, resulting in an increased WF[14] (**Figure *3*f**).

Topography and CPD maps obtained from KPFM for *poly-ITO-µm-grain* films are presented in **Figure *3*b-c**. Using the dash-framed grains and arrows as guiding marks, we note that the indicated grains exhibit a lighter coloration, implying a higher applied CPD and thus a lower WF. The respective grains in the EBSD map (out-of-plane, z-direction) are coloured green, indicating (111) orientation (**Figure *3*d-e, S6**). Conversely, the darker-coloured grains identified on the CPD map, reflecting a low applied CPD, thus a high WF, align with the red/blue coloured grains on the EBSD map, which correspond to the (001) family of planes. The preferred surface termination of a single grain along 4×(001) was also confirmed with high-resolution transmission electron microscopy (HR-TEM) and selected area diffraction pattern (SADP) as presented in **Figure 3g-h**.

Furthermore, we conducted EBSD on *poly-ITO-nm-grain* films, as shown in **Figure *3*i-j and Figure S7**. The distinct crystal orientations of the nm-scale grains are visible, suggesting that local WF variations induced by the grain's orientation could also be expected. However, due to the nm-scale of the grains, the spatial resolution of the KPFM tip was not sufficient to detect such nanoscale variation. Despite this, it is suggested that local WF variations, even at the nanoscale, are expected on any polycrystalline sample[47, 51]. The presence of significant energetic variations across a material can lead to unwanted effects on a device level, such as an uneven charge distribution, altered electronic transport properties, and even limitations in device efficiency and performance as previously reported[37, 38, 52].

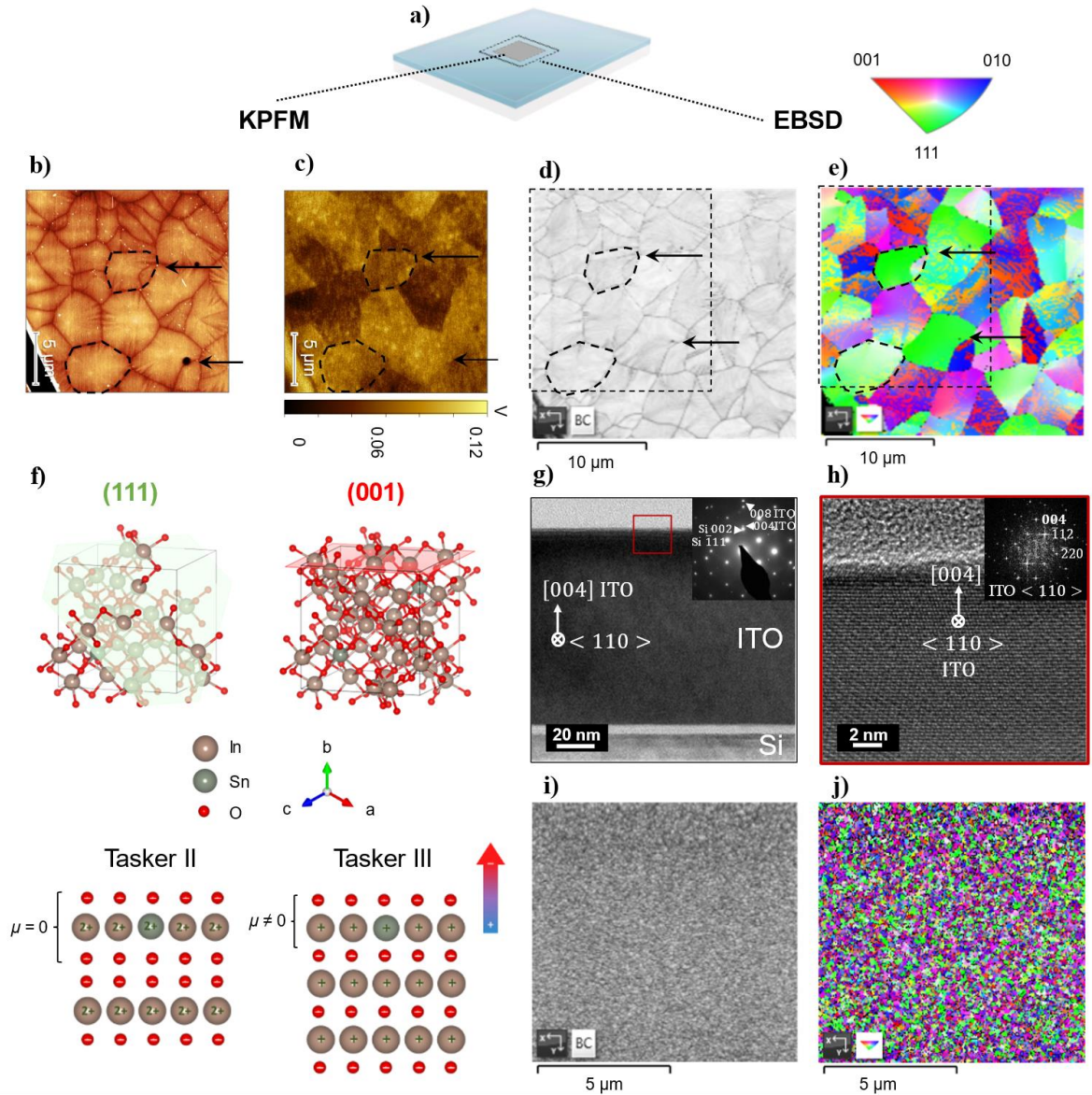

**Figure 3.** Grain orientation influence on local WF. **(a)** Schematic representation of the point of interest (POI) on the sample; **(b)** Topography and **(c)** CPD mapping from KPFM measurement; **(d)** SEM image and **(e)** EBSD mapping of *poly-ITO-µm-grain* ITO film; **(f)** representation of (001) and (111) plane orientation for ITO and analogous Tasker surfaces; **(g)** Cross-sectional TEM image and SADP (inset) of the *poly-ITO-µm-grain* film and **(h)** HR-TEM and corresponding fast-Fourier-transform (FFT) (inset) image of the top surface showing preferential surface termination in 4x(001) orientation; **(i)** SEM image and **(j)** EBSD mapping of the *poly-ITO-nm grain* ITO film (Black dashed square in **d-e** represents the area on which both KPFM and EBSD were performed, arrows and dash-framed grains serve as guiding marks).

To verify the WF values determined by KPFM and to extract the valence band maximum (VBM) and the highest occupied molecular orbital (HOMO) levels, UPS measurements were conducted. **Figure *4*a** showcases the WF values determined by KPFM for the studied ITOs, in their initial state (right upon solution cleaning procedure), after UV-$O_3$ treatment, and subsequent 2PACz-SAMs deposition. The large error bar for the *poly-ITO-µm-grain* film, could be ascribed to the presence of the distinct crystalline grain orientations. However, the overall WF of all bare ITOs are found within the same range, and a systematic increase in the WF after UV-$O_3$ treatment and upon 2PACz-SAMs anchoring is evident across all studied ITO films. This trend is expected, as any form of surface treatment inherently alters the surface potential and thereby directly impacts the WF[47].

UPS results complement these findings, indicating a pronounced shift of the secondary electron cut-off (SECO) towards lower binding energies. This shift strongly implies a substantial WF increase upon the anchoring of 2PACz-SAMs, a phenomenon consistent across all the examined cases. In addition, the VBM region spectra show a significant modification after 2PACz-SAMs deposition. Specifically, the characteristic sharp, linear-like form typical of TCOs – attributed to localized electronic states changes to a hump-like HOMO edge, characteristic of organic molecules with delocalized π-electrons[53] (**Figure *4*b**). The presence of 2PACz-SAMs on the surface was further confirmed via X-ray photoelectron spectroscopy (XPS) analysis (**Figure S8-9**).

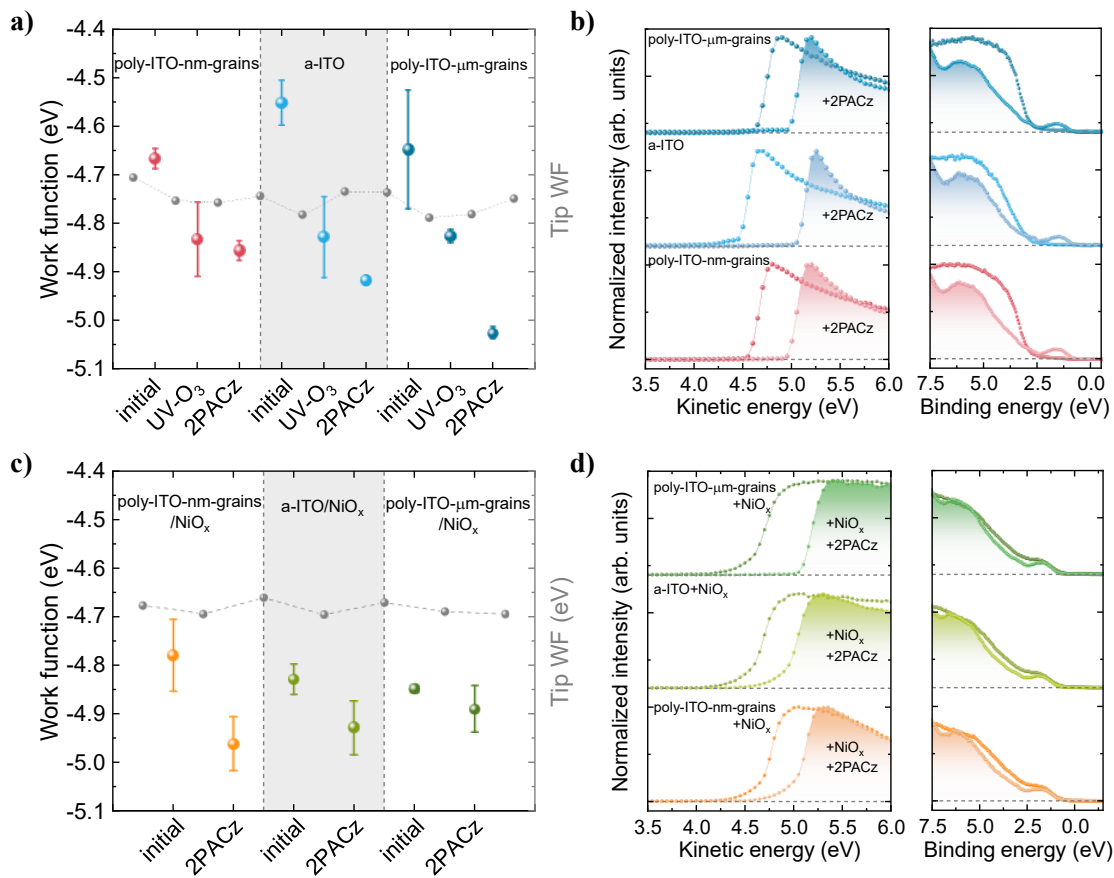

**Figure 4**. KPFM WF values **(a, c)**, UPS secondary electron cut-off (SECO) and valence band region **(b, d)** for ITO with and without 2PACz **(a, b)**; and ITO/NiO$_x$ with and without 2PACz **(c-d)**. (The tip WF is presented in (**a, c**) for reference).

As observed in **Figure 2b** (left column), the introduction of an amorphous NiO$_x$ layer appears to mitigate surface potential disparities arising from the distinct microstructure of the ITO films. Measuring the average WF values by KPFM and UPS, only a slight variation in WF is observed for the different ITOs after NiO$_x$ and NiO$_x$/2PACz-SAMs deposition (**Figure 4c-d**).

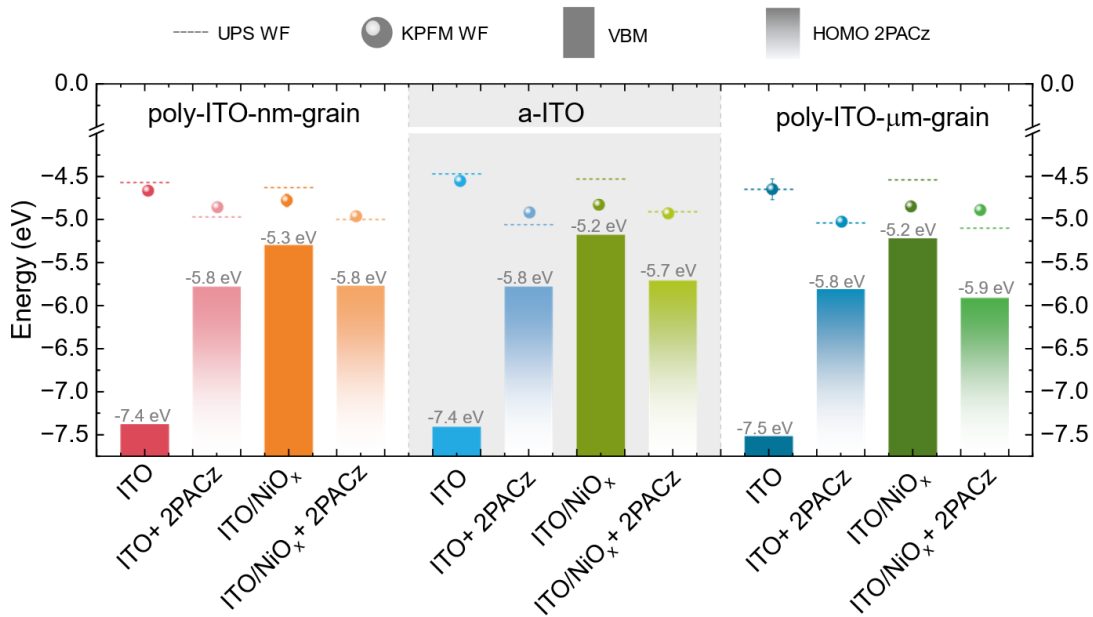

**Figure 5**. Overview of UPS and KPFM data and reconstructed energy level diagram for the distinct stacks studied (spheres and dashed lines represent work function values determined by KPFM and UPS, respectively; grey numbers refer to ionization potential, $I_p$, which corresponds to VBM or HOMO values for TCOs and TCO/2PACz stack, respectively).

The reconstructed energy level diagram represented in **Figure 5**, shows the values of the work function determined by KPFM (sphere) and UPS (dashed line) and ionization potential, $I_p$ (numerical values in grey font). All values are listed in the **Table S2**. An overall good correlation between two independent techniques (each operating at different conditions, i.e., UPS at UHV and KPFM at ambient air) and a consistent trend in the WF increase upon ITOs surface modification can be observed (for a detailed explanation, check **Figure S10**).

As expected, all three ITO films have pronounced n-type character as their Fermi level is far from VBM. Modified ITOs, either with 2PACz-SAMs or $NiO_x$ only, or a combination $NiO_x$/2PACz-SAMs shifted the Fermi level closer to the ionization potential level (HOMO or VBM, respectively), indicating the p-type characteristics and enhanced hole selectivity. It is worth highlighting that the values of the WF and VBM for ITO/NiO$_x$ agree with previously reported values for ALD deposited NiO$_x$ (4.6 – 4.7 eV and 5.3 eV, respectively)[37]. Upon depositing 2PACz-SAMs, the WF is further increased, due to the molecular dipole moment that 2PACz-SAMs introduce[44], contributing to the surface energy term[47]. Notably, the WF values for the ITO/2PACz-SAMs case and the ITO/NiO$_x$/2PACz-SAMs case are quite similar. Going back to the KPFM mapping, we suggest that the main advantage of NiO$_x$ in addition to its hole transport properties is its amorphous nature, minimizing pin hole formation and ensuring a uniform WF distribution due to the lack of preferential grain orientation. We therefore argue that the use of an amorphous metal oxide with adequate WF for hole (or electron) extraction, or an amorphous TCO buffer layer ensures an enhanced coverage of the 2PACz-SAMs and with it, a uniform WF distribution. This combination holds the potential to improve energy level alignment and charge extraction on a device scale, thereby enhancing overall device performance, reproducibility, and stability.

**Conclusion**

In summary, we studied the impact of the ITO microstructure on the electronic properties of hole-selective transport layers, NiO$_x$ and 2PACz-SAMs. Three different types of ITO thin-films morphology and microstructure were characterized. Correlated KPFM and EBSD mapping revealed that polycrystalline ITO films present a non-uniform distribution of surface potential, subsequently impacting WF uniformity. The application of 2PACz-SAMs was not sufficient to overcome the lateral inhomogeneity in WF inherent to the polycrystalline ITO films. However, this challenge can be successfully addressed by employing either an amorphous TCO or an amorphous NiO$_x$ buffer layer. While polycrystalline TCOs present the potential for high mobility, wide band gap and low visible to infrared absorption, it is relevant to look for strategies, such as an amorphous buffer layer, to ensure a uniform WF potential distribution.

ASSOCIATED CONTENT

**Supporting Information**

Detailed description of used materials, applied experimental methods and procedures; elaborated XRD data, annealing series data, Tauc plot, KPFM and UPS WF values in tabular format, KPFM maps of ITO/NiO$_x$ KCl treated surface, XPS survey spectra, EBSD analysis.

AUTHOR INFORMATION

**Corresponding authors**

**Suzana Kralj** – MESA+ Institute for Nanotechnology, University of Twente, Enschede 7500 AE, The Netherlands; orcid.org/0000-0003-2847-8359; E-mail: s.k.kralj@utwente.nl

**Monica Morales-Masis** – MESA+ Institute for Nanotechnology, University of Twente, Enschede 7500 AE, The Netherlands; orcid.org/0000-0003-0390-6839; E-mail: m.moralesmasis@utwente.nl

**Authors**

**Pia Dally** – KAUST Solar Center (KSC), Physical Sciences and Engineering Division (PSE), King Abdullah University of Science and Technology (KAUST), Thuwal 23955-6900, Kingdom of Saudi Arabia;

**Pantelis Bampoulis** – Physics of Interfaces and Nanomaterials, MESA+ Institute for Nanotechnology, University of Twente, Enschede 7500 AE, The Netherlands; orcid.org/0000-0002-2347-5223

**Badri Vishal** – KAUST Solar Center (KSC), Physical Sciences and Engineering Division (PSE), King Abdullah University of Science and Technology (KAUST), Thuwal 23955-6900, Kingdom of Saudi Arabia; orcid.org/0000-0002-3480-6841

**Stefaan De Wolf** – KAUST Solar Center (KSC), Physical Sciences and Engineering Division (PSE), King Abdullah University of Science and Technology (KAUST), Thuwal 23955-6900, Kingdom of Saudi Arabia; orcid.org/0000-0003-1619-9061

**Notes**

The authors declare no competing financial interest.

ACKNOWLEDGMENT

This project was supported by the Solar-ERA.NET CUSTCO project by Netherlands Enterprise Agency (RVO under contract SOL18001 and by the European Research Council (ERC) under the European Union's Horizon 2020 Research and Innovation Program (CREATE, Grant Agreement No. 852722).

The authors would like to acknowledge P-A. Repecaud and Y. Smirnov for input information on PLD depositions, M. Smithers for SEM and EBSD imaging, D. M. Cunha, R. Saive, L. Chen and

**Supporting information**

# Impact of TCO Microstructure on the Electronic Properties of Carbazole-based Self-Assembled Monolayers

*Suzana Kralj [1a]\*, Pia Dally [2], Pantelis Bampoulis [1b], Badri Vishal [2], Stefaan De Wolf [2], Monica Morales-Masis [1a]\**

[1a] S. Kralj, M. Morales-Masis

MESA+ Institute for Nanotechnology, University of Twente, Enschede 7500 AE, The Netherlands

[1b] P. Bampoulis

Physics of Interfaces and Nanomaterials, MESA+ Institute for Nanotechnology, University of Twente, Enschede 7500 AE, The Netherlands

[2] P. Dally, B. Vishal, S. De Wolf

KAUST Solar Center (KSC), Physical Sciences and Engineering Division (PSE), King Abdullah University of Science and Technology (KAUST), Thuwal 23955-6900, Kingdom of Saudi Arabia

**Corresponding Author**

\* S. Kralj (s.k.kralj@utwente.nl), \* M. Morales-Masis (m.moralesmasis@utwente.nl)

**Methods Section**

**Materials**

ITO commercial substrates were purchased from *Ossila Ltd*. SnO$_2$/In$_2$O$_3$ (10/90 wt%) target from *Toshima Materials Co., Ltd*, 2-(9H-Carbazol-9-yl)ethyl)phosphonic acid (2PACz), >98.0% from *Tokyo Chemical Industry Co., Ltd*, Anhydrous Ethanol (max. 0.003% H$_2$O, ≥99.8%) from *VWR Chemicals*, potassium chloride (KCl), 99.0% from *ThermoFisher*.

**Methods**

Full area glass/ITO substrates (15 mm x 20 mm) were cleaned through sonication in Hellmanex® solution (2 vol.% in deionized water), deionized water, acetone and 2-propanol consecutively for 10 minutes each and dried using nitrogen gun. SAMs solution was prepared following the procedure reported by Al-Ashouri *et al*.[1] In short, 1 mg/mL 2PACz was dissolved in anhydrous ethanol. Prior to spin-coating, glass/ITO substrates were treated 15 min in UV-O$_3$ to activate the surface. 100 μL 2PACz solution was then spin-coated statically on samples with freshly activated surface. After approximately 5 s resting of 2PACz solution, spin-coating program was started. Used settings were 5000 rpm for 30 s. Upon that, annealing step was performed at 100 °C for 10 minutes to allow phosphonic acid anchoring group to bind to the substrate. To remove any excess unbound molecules, washing step was introduced. 2 x 100 μL of anhydrous ethanol was spin-coated dynamically using the same program. As a final step, 5 min drying at 100 °C was done.

***Pulsed Laser Deposition of ITO thin films.*** Indium tin oxide (ITO) films grown in house (University of Twente) were prepared by Pulsed Laser Deposition (PLD) technique following the procedure previously reported by our group[2]. Films were deposited on glass substrates (*ECOGlass SCHOTT*) which were cleaned prior to deposition in ultrasonic bath for 5 minutes in acetone and isopropanol and rinsed in deionised water, respectively. All depositions were performed in wafer-scale PLD system designed by *Solmates BV*. $SnO_2/In_2O_3$ (90/10 wt%) target was ablated using Compexpro (*COHERENT*) KrF excimer laser ($\lambda = 248$ nm) with a fluence of 2.6 J/cm$^2$ and frequency of 10 Hz. Deposition pressure was controlled to 0.02 mbar (50/50% Ar/O$_2$). Number of scans (pulses) was adjusted to keep the thickness of as deposited film at approximately 100 nm. Exact thickness value was estimated by X-Ray Reflectivity (XRR).

In order to further improve optical and electrical properties, additional annealing step was performed on as-deposited PLD ITO films. Annealing was performed under controlled environment of 5% hydrogen in nitrogen gas atmosphere. Temperature was set to 450 ˚C for 20 minutes.

***Deposition of NiO$_x$ hole transport layer.*** NiO$_x$ layers were deposited by radio frequency (RF, 13.56 MHz) magnetron sputtering from NiO stoichiometric target (99.95% purity) at room temperature using *Angstrom Engineering EvoVac* system at KAUST Solar Centre. The base pressure for deposition was <$5\times10^{-7}$ Torr. Deposition was performed under 3 mTorr with 20 sccm Argon gas flow. Before each deposition, to remove the contamination layer on the surface of the target (if any), the target was pre-sputtered for 15 min. For homogenous deposition, substrate rotation was provided. Thickness was controlled to ~14 nm (measured by ellipsometry on co-deposited silicon substrate).

***KCl surface passivation.*** KCl was dissolved in deionized water (3.5 mg/mL). Solution was put on vortex mixer for 30 minutes. 120 μL solution was spin-coated on NiOx film followed by 10 minutes annealing at 130 ˚C. For case of further processing 2PACz layer, no additional surface activation step was performed and the spin-coating setting are the same as for ITO substrates.

***Kelvin Probe Force Microscopy (KPFM).*** Amplitude modulated Kelvin Probe Force Microscopy (AM-KPFM) was measured by Bruker Dimension Icon AFM using SCM-PIT-v2 tip in dual pass mode. All measurements were performed under ambient conditions. Calibration of the tip was executed using reference samples: i) PFKPFM-SMPL (Kelvin Probe Sample Aluminium and Gold on Silicon) at the beginning of measurement, before changing the sample and at the end of measurements to track stability of probe and ii) freshly cleaved highly ordered pyrolytic graphite (HOPG) surface before each measurement to accurately calibrate probe. Area of 10x10 μm$^2$ was scanned in all cases. The mapping was performed with scanning speed of 0.75 Hz. Scanning resolution was 512x512 lines in fast and slow-scan axis, respectively. For the second pass, z-height of 10 nm was used. For each condition minimum of 3 (or more) spots on the sample were measured to ensure enough data set for statistical analysis. The work function was calculated using formula[3, 4]:

$$CPD = \frac{\phi_{Probe} - \phi_{Sample}}{|e|}$$

Which is modified for probe calibration to:

$$\phi_{Probe} = \phi_{Reference} + CPD \cdot |e|$$

while for sample work function calculation to:

$$\phi_{Sample} = \phi_{Probe} - CPD \cdot |e|$$

where *CPD* equals to applied average contact potential difference, *Φ$_{Probe}$* represents the work function of probe (SCM-PIT-v2), *Φ$_{Refernce}$* represents the work function of reference samples (theoretical work function values used for reference samples were -5.1 eV and -4.5 eV for Au and HOPG[3], respectively) and *Φ$_{Refernce}$* represents the work function of the sample.

***Ultraviolet-Visible-Near Infrared Spectrophotometry (UV-Vis-NIR).*** Optical properties of ITO thin-films were measured with *PerkinElmer Lambda 950S* UV-Vis-NIR spectrophotometer with integrating sphere. Transmittance (T) and reflectance (R) were measured while absorptance (A) was calculated as: A = 100 – T – R.

*Hall effect.* Electrical properties were measured using a Hall effect measurement with *ezHEMS NanoMagnetics Instruments* set-up in Van der Pauw configuration at the room temperature and under magnetic field, $B = 0.9642$ T. Samples were cut into 10 mm x 10 mm pieces, input thickness value was 100 nm and applied source current, $I = 100$ μA.

*Corelation KPFM and EBSD measurement.* The protocol procedure by Maryon et al. was followed[4]. Initially, the sample's surface was mechanically scratched to establish a reference mark for subsequent point-of-interest (POI) identification. Subsequently, solution cleaning procedure was performed to remove any residual particles from the surface. The reference mark was firstly found with optical microscope and it was marked as a point of origin. KPFM measurements were conducted on randomly selected area close to proximity of reference line and its coordinates (X, Y) were stored for later co-localized measurement. The parameters used are the same as for standard KPFM measurements. EBSD was performed after KPFM measurements to prevent deposition of unwanted carbon coating on the sample surface due to exposure to the electron beam. Sample was mounted on a SEM holder using silver paste to ensure good contact and conductivity of the top surface to prevent static electric charge accumulation. The sample-carrying stage was tilted to 70˚ to enhance the diffracted signal. The scanned surface was adjusted to be within the same region as mapped with KPFM. Measurements were conducted with the 15 kV electron acceleration voltage and at working distance < 15 mm.

*Transmission electron microscopy (TEM) and focused ion beam (FIB).* For the Transmission Electron Microscopy (TEM)-based study, a cross-sectional electron-transparent lamella was meticulously prepared using a focused ion beam (FIB) within the Scanning Electron Microscope (SEM-FIB Helios G5 DualBeam, FEI), equipped with an EasyLift nanomanipulator and a Gallium (Ga) ion source. To protect the region of interest during FIB processing, a multi-layer protection layer were deposited. Initially, a 0.5 µm layer of both Carbon (C) and Tungsten (W) was deposited using an electron beam (e-beam), followed by an additional 3 µm layer of W deposited by the ion beam, providing a robust protection to thin film. The ion beam milling proedure was carried out step by step with decresing beam currents (ranging from 2.4 nA to 0.025 nA, over an accelerating voltage range of 30 kV to 5 kV). This precise process allowed for the gradual cutting and thinning of the lamella down to a thickness of 60 nm while minimizing ion beam-induced damage. Furthermore, a low-current cleaning procedure (performed at voltages ranging from 5 kV to 2 kV and currents from 81 pA to 28 pA) was executed to ensure the removal of any potential contamination. Subsequent TEM-based experiments were conducted using the Cs-corrected ThermoFisher Titan 60-300 Cubed TEM microscope, operating at an acceleration voltage of 300 kV. The acquired TEM data underwent thorough processing using specialized software packages, including Gatan™ Digital Micrograph and Thermo Scientific™ Velox suites.

***Structure visualization Software.*** *ChemDraw* molecule editor and visualization software was used to create molecule of 2PACz. *VESTA* was used to visualize plane orientations in $In_2O_3$. Input data file was taken from Crystallography Open Database (COD) for $In_2O_3$ (*CIF 2310009*)[5] and modified with replacing two Indium (In) atoms for Tin (Sn) atoms to illustrate the Sn-doping of $In_2O_3$.

***UPS/XPS***. Ultraviolet photoelectron spectroscopy (UPS) and X-ray photoelectron spectroscopy (XPS) measurements were conducted using an Omicron multi-probe Photoemission Spectroscopy (PES) system (base pressure of $5 \times 10^{-10}$ mbar) equipped with a Sphera II EAC 125 hemispherical electron analyser and a multi-channeltron electron detector. The surface WF and valence region were studied by UPS with a vacuum ultraviolet unfiltered He(1) (21.22 eV) source (focus) (90 mA, 570 V). The samples were biased to -10 V to observe the secondary electron cut-off (SECO)[6]. The photoelectrons were collected at an angle of 80° between the sample and analyser, with a normal electron take off angle. The constant analyser pass energy (CAE) was 5 eV for the valence band region and for the SECO.

XPS was conducted with a monochromatic Al Kα 1486.6 eV X-ray source, operating at a power of 390 W. The photoelectrons were also measured from the same spot as used for UPS. The photoelectron pass energy was fixed at 15 eV for high-resolution XPS and 50 eV for survey. Quantification of XPS spectra were conducted in CASAXPS software, integrating the peak areas using a Tougaard based background function.

The value of ionization potential, $I_P$ was calculated from UPS data using formula[7]:

$$I_p = 21.2\ eV - (E_{SECO} - E_{VB\ or\ HOMO, UPS})$$

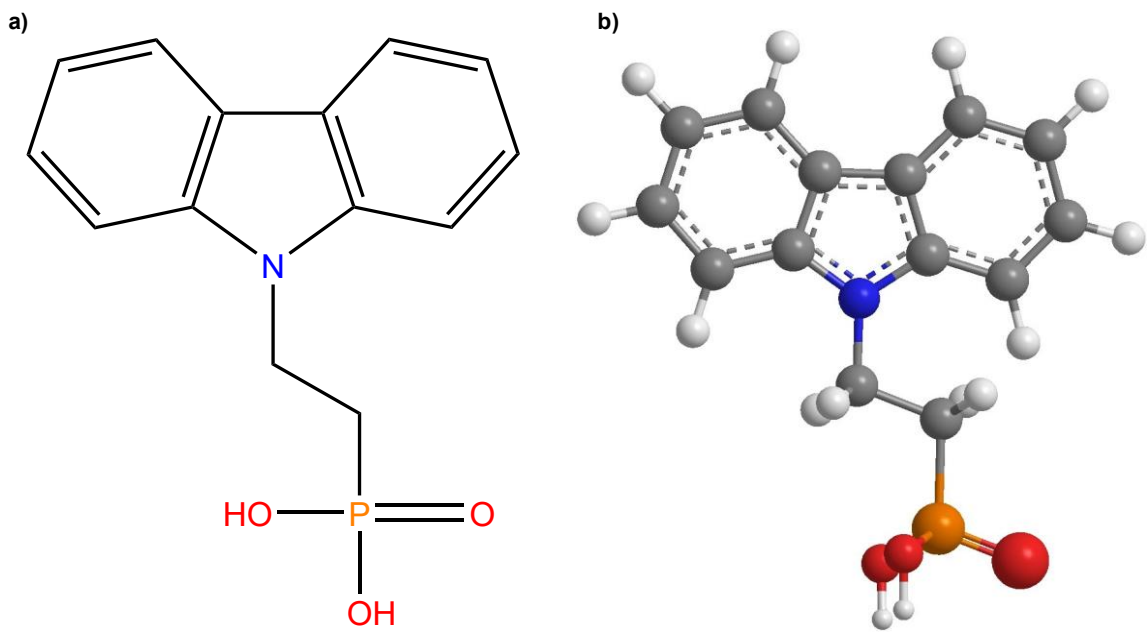

**Figure S1.** [2-(9H-carbazol-9-yl)ethyl] phosphonic acid (2PACz): **a)** Chemical structural formula and, **b)** Three-dimensional structure (atom coloration: carbon (C) in grey, hydrogen (H) in white, nitrogen (N) in blue, phosphorous (P) in orange, and oxygen (O) in red.).

**Table S1.** X-Ray diffraction peak positions for polycrystalline samples

| $In_2O_3$ reference (ICSD code: 14388) | | Commercial ITO (polycrystalline, nm-size grains) | PLD ITO anneal (polycrystalline, μm-size grains) |
|---|---|---|---|
| (hkl) | 2Theta | 2Theta (°) | 2Theta (°) |
| (211) | 21.50 | 21.23 | 21.51 |
| (222) | 30.59 | 30.22 | 30.75 |
| (400) | 35.46 | 35.09 | 35.62 |
| (411) | 37.69 | 37.28 | 37.72 |
| (422) | 43.80 | / | 43.86 |
| (341) | 45.69 | 45.22 | 45.82 |
| (440) | 51.02 | 50.42 | 51.25 |
| (611) | 55.98 | / | 56.20 |

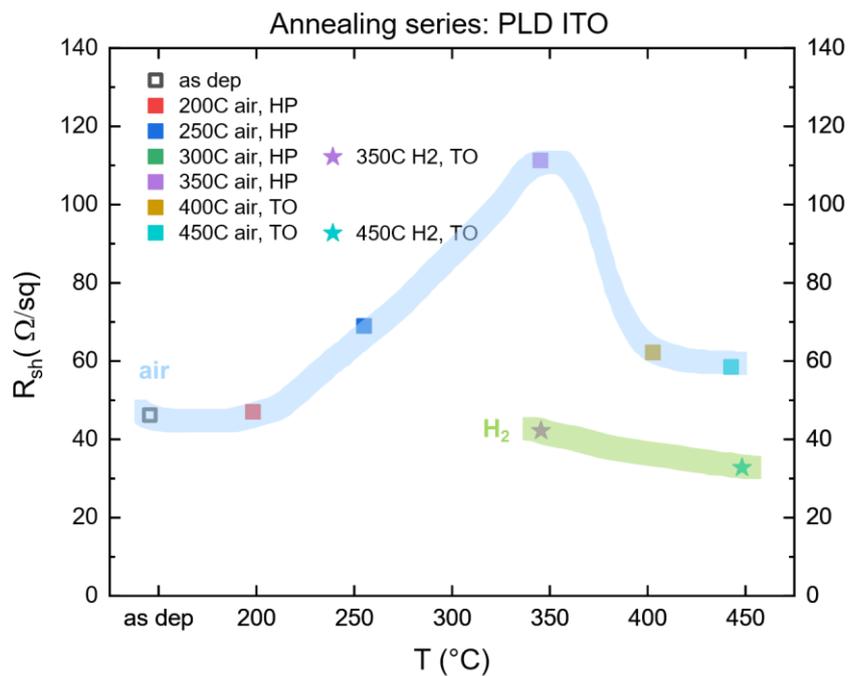

**Figure S2.** Post-annealing treatment of PLD ITO films at different temperatures and under different atmospheres showing sheet resistance value upon annealing, displaying sheet resistance values upon annealing. The film annealed at 450 °C under a 5% $H_2/N_2$ atmosphere exhibited the lowest sheet resistance value and was consequently chosen for further investigation.

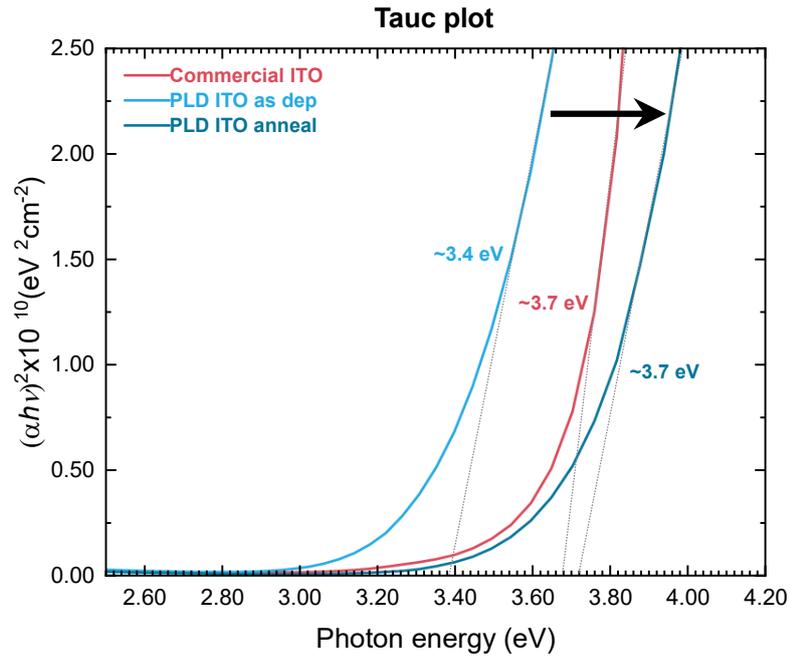

**Figure S3.** Tauc plot for studied ITO films with extracted band gap value

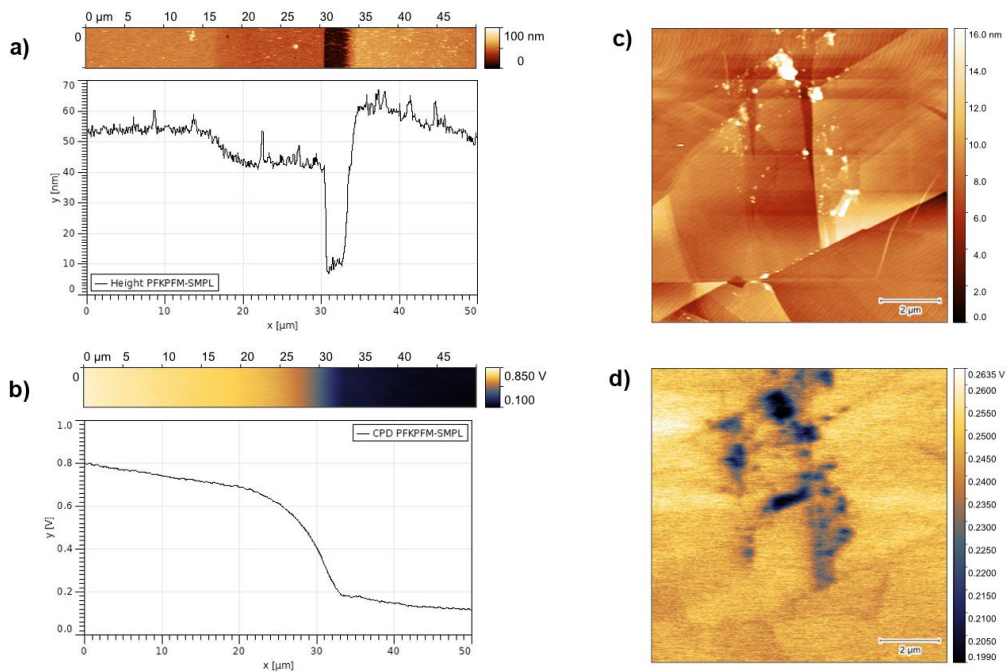

**Figure S4.** KPFM probe calibration (SCM-PIT v2 ). Topography (**a,c**) and Contact potential difference mapping (**b,d**) for PFKPFM-SML (**a,b**) and HOPG (**c,d**) reference samples.

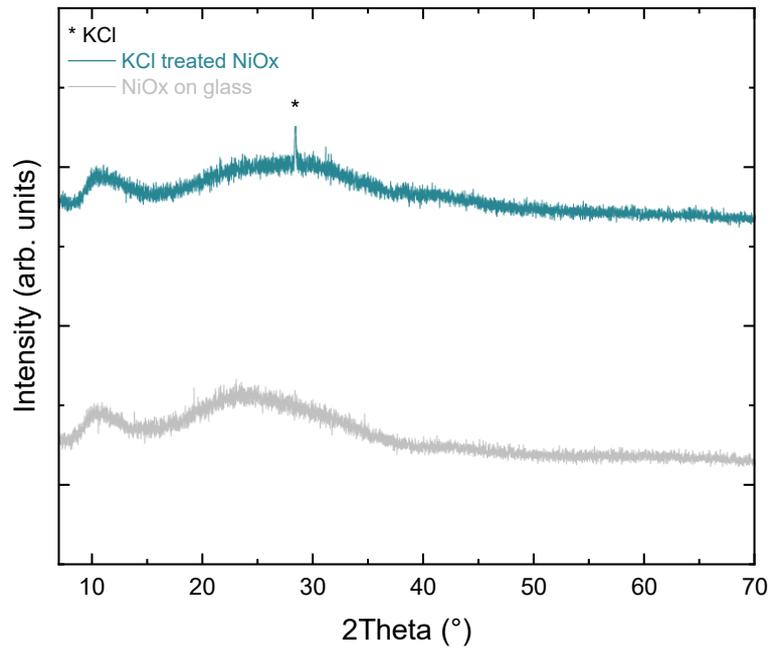

**Figure 5.** XRD of NiO$_x$ (grey) and KCl treated NiO$_x$ film (blue) indicating the presence of KCl on the surface.

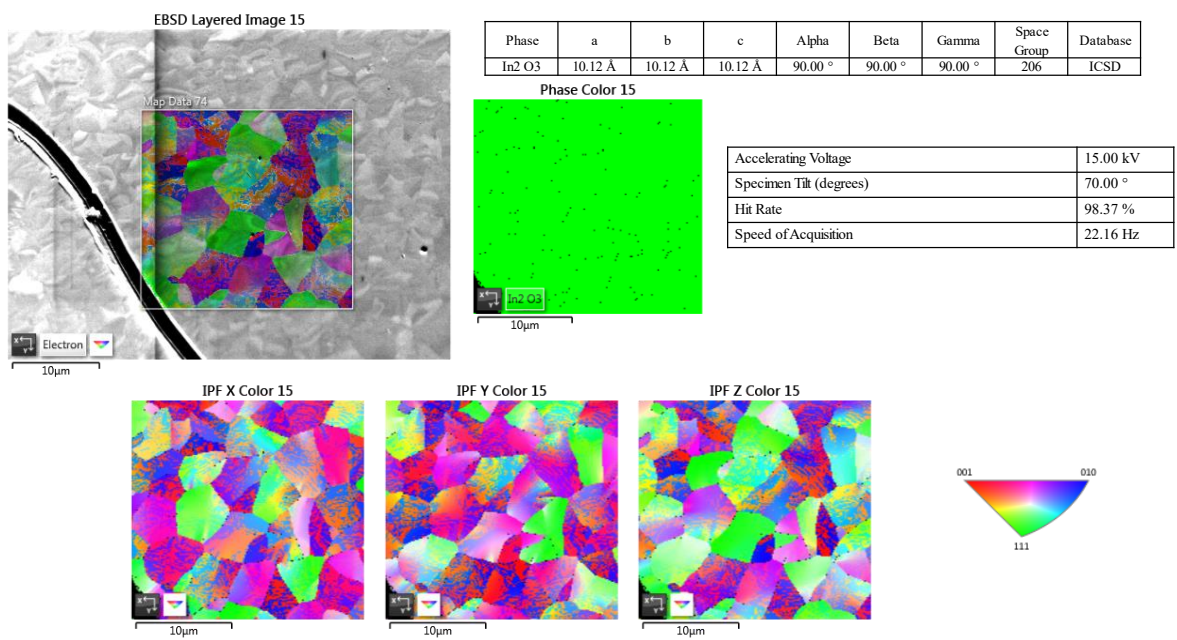

**Figure S6**. Comprehensive EBSD characterization of poly-ITO-µm-grain film.

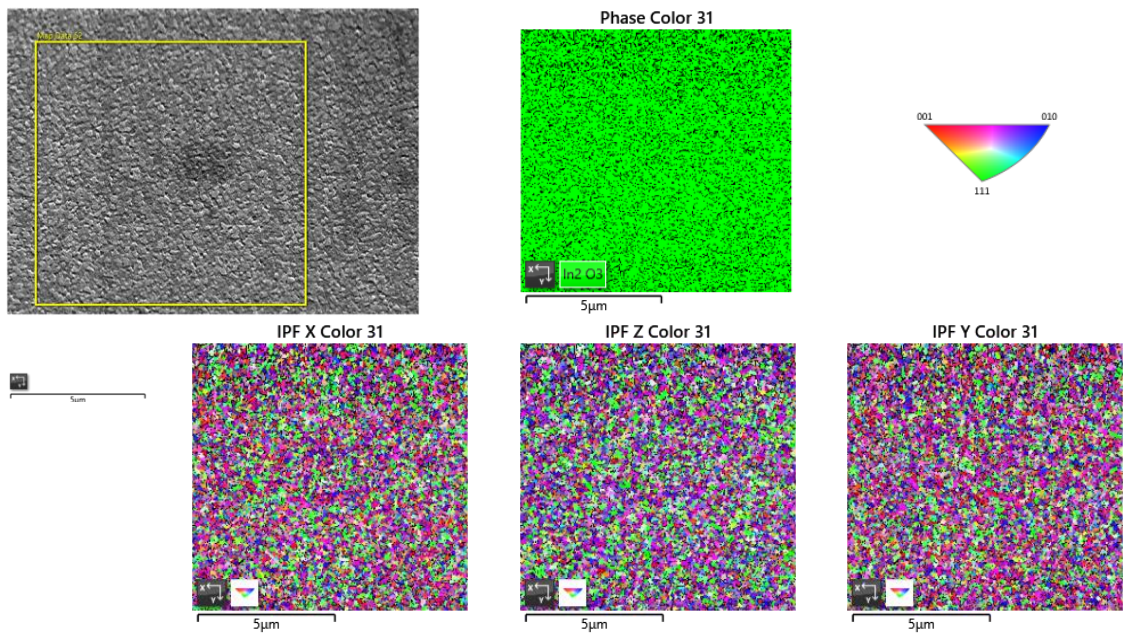

**Figure S7.** Comprehensive EBSD characterization of poly-ITO-nm-grains film.

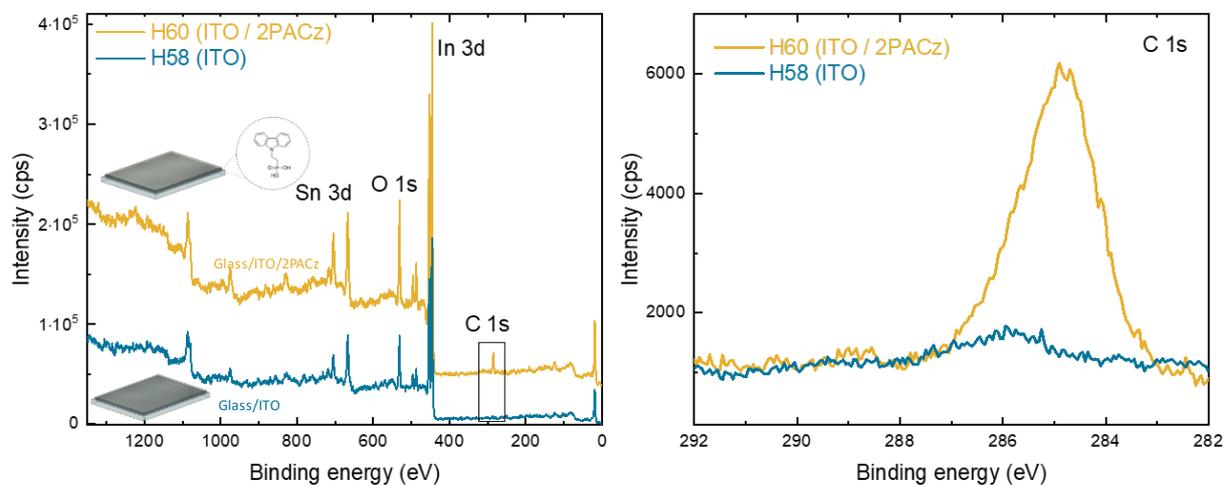

**Figure S8.** XPS survey spectra for ITO (blue) and ITO + 2PACz (yellow). The high resolution XPS spectra for C 1s is shown, indicating the presence of 2PACz-SAM

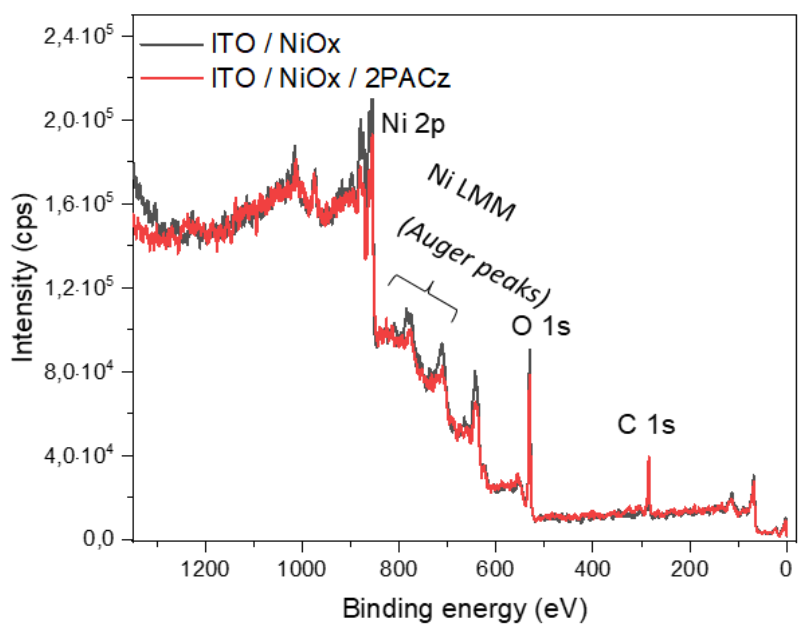

**Figure S9.** XPS survey spectra for ITO/NiO$_x$ (black) and ITO/NiO$_x$ + 2PACz (red)

**Table S2.** Summary of average work function values determined by UPS and KPFM

| Condition | WF UPS (eV) | WF KPFM (eV) | Energy onset, $E_{VB\ or\ HOMO,\ UPS}$ (eV) | Ionization potential, $I_p$ (eV) |
|---|---|---|---|---|
| Commercial ITO | -4.57 | -4.66 | 2.81 | -7.38 |
| Commercial ITO UV-O$_3$ | | -4.83 | | |
| Commercial ITO + 2PACz | -4.97 | -4.86 | 0.81 | -5.78 |
| Commercial ITO / NiO$_x$ | -4.63 | -4.77 | 0.67 | -5.30 |
| Commercial ITO / NiO$_x$ + 2PACz | -5.00 | -4.96 | 0.77 | -5.77 |
| PLD ITO as dep | -4.47 | -4.55 | 2.94 | -7.41 |
| PLD ITO as dep UV-O$_3$ | | -4.83 | | |
| PLD ITO as dep + 2PACz | -5.06 | -4.92 | 0.72 | -5.78 |
| PLD ITO as dep / NiO$_x$ | -4.53 | -4.83 | 0.65 | -5.18 |
| PLD ITO as dep / NiO$_x$ + 2PACz | -4.91 | -4.93 | 0.80 | -5.71 |
| PLD ITO anneal | -4.65 | -4.65 | 2.87 | -7.52 |
| PLD ITO anneal UV-O$_3$ | | -4.83 | | |
| PLD ITO anneal + 2PACz | -5.04 | -5.02 | 0.77 | -5.81 |
| PLD ITO anneal / NiO$_x$ | -4.54 | -4.85 | 0.68 | -5.22 |
| PLD ITO anneal / NiO$_x$ + 2PACz | -5.10 | -4.89 | 0.81 | -5.91 |

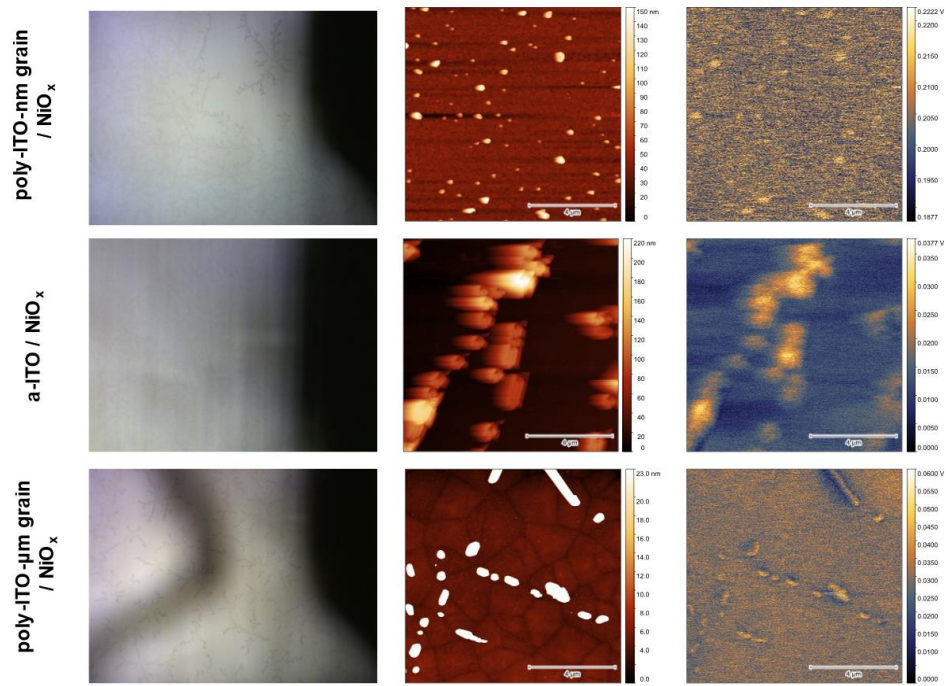

**Figure S10.** ITO/NiO$_x$ films - areas with KCl crystals. Left column showcases KCl crystals on ITOs surface captured by optical microscope of AFM setup, middle column displays topography scan, and right column presents surface potential distribution scans showing domains of KCl crystals with higher surface potential

A slight discrepancy between KPFM-UPS WF values were observed for different ITO samples modified with NiO$_x$ and subjected to surface passivation treatment using a KCl solution. This can be attributed to the recrystallization of KCl on the surface of NiO$_x$ and the non-uniform distribution of KCl crystals, as previously reported by Zheng et al.[8]. The KPFM mapping presented in **Figure 2.b**, left-hand column are selected to show areas without KCl crystals, while **Figure S8**. shows the areas with KCl crystals and their effect on surface potential distribution on the ITO/NiO$_x$ stack. To further confirm that the observed particles are indeed KCl crystals, XRD was performed. From **Figure S5**. it can be seen that as deposited NiO$_x$ film on glass is amorphous, while upon KCl treatment, clear peak is observed at 28.42° 2Theta which corresponds to (200) plane of KCl. On the other side, UPS spot size is in the order of mm.

Therefore, these crystals were also included in averaging out the values of Fermi level causing differences in values. Nonetheless, upon depositing 2PACz, the differences in WF values decrease again as the KCl crystals are washed off by solution when processing SAMs on top.